\documentclass[lettersize,journal]{IEEEtran}
\usepackage{amsmath,amsfonts,amssymb}
\usepackage{graphicx}
\usepackage{subfigure}
\usepackage{algorithmicx}
\usepackage{algorithm}
\usepackage{algpseudocode}
\usepackage{textcomp}
\usepackage{makecell}
\usepackage{multirow,multicol}
\usepackage[colorlinks]{hyperref}
\usepackage[table]{xcolor}
\usepackage{threeparttable,flushend}

\newcommand*\LyXZeroWidthSpace{\hspace{0pt}}

\DeclareFontEncoding{LGR}{}{}
\DeclareTextSymbol{\~}{LGR}{126}
\floatstyle{ruled}
\newfloat{algorithm}{tbp}{loa}
\providecommand{\algorithmname}{Algorithm}
\floatname{algorithm}{\protect\algorithmname}

\begin{document}

\title{A First Look at the Performance Enhancement Potential of Fluid Reconfigurable Intelligent Surface}

\author{Abdelhamid Salem,~\IEEEmembership{Member,~IEEE}, 
            Kai-Kit Wong,~\IEEEmembership{Fellow,~IEEE},\\
            George Alexandropoulos,~\IEEEmembership{Senior~Member,~IEEE}, 
            Chan-Byoung Chae,~\IEEEmembership{Fellow,~IEEE}, and\\ 
            Ross Murch,~\IEEEmembership{Fellow}, \textit{IEEE}
\vspace{-5mm}

\thanks{The work of A. Salem and K.-K. Wong was supported by the Engineering and Physical Sciences Research Council (EPSRC) under Grant EP/W026813/1. The work of C.-B. Chae This work was also
supported by IITP (IITP-2025-RS-2024-00428780, IITP-2021-0-00486) grants funded by the Korean government. The work of R. Murch was supported by the Hong Kong Research Grants Council Area of Excellence grant AoE/E-601/22-R.}
\thanks{A. salem and K.-K. Wong are with the Department of Electronic and Electrical Engineering, University College London, London, United Kingdom. K.-K. Wong is also with Yonsei Frontier Lab, Yonsei University, Seoul, Korea (e-mails: \{a.salem, kai-kit.wong\}@ucl.ac.uk). A. Salem is also affiliated with Benghazi University, Benghazi, Libya.}
\thanks{G. C. Alexandropoulos is with the Department of Informatics and Telecommunications, National and Kapodistrian University of Athens, 16122 Athens, Greece and the Department of Electrical and Computer Engineering, University of Illinois Chicago, IL 60601, USA (e-mail: alexandg@di.uoa.gr)}
\thanks{C.-B. Chae is with the School of Integrated Technology, Yonsei University, Seoul, 03722, Korea (e-mail: cbchae@yonsei.ac.kr).}
\thanks{R. Murch is with the Department of Electronic and Computer Engineering and Institute for Advanced Study (IAS), The Hong Kong University of Science and Technology, Clear Water Bay, Hong Kong SAR, China (e-mail: eermurch@ust.hk).}
\thanks{Corresponding author: Kai-Kit Wong.}
}
\maketitle

\begin{abstract}
The fluid antenna concept represents shape-flexible and position-flexible antenna technologies designed to enhance wireless communication applications. In this paper, we apply this concept to reconfigurable intelligent surfaces (RISs), introducing fluid RIS (FRIS), where each tunably reflecting element becomes a {\em fluid element} with additional position reconfigurability. This new paradigm is referred to as fluid RIS (FRIS). We investigate an FRIS-programmable wireless channel, where the fluid metasurface is divided into non-overlapping subareas, each acting as a fluid element that can dynamically adjust both its position and phase shift of the reflected signal. We first analyze the single-user, single-input single-output (SU-SISO) channel, in which a single-antenna transmitter communicates with a single-antenna receiver via an FRIS. The achievable rate is maximized by optimizing the fluid elements using a particle swarm optimization (PSO)-based approach. Next, we extend our analysis to the multi-user, multiple-input single-output (MU-MISO) case, where a multi-antenna base station (BS) transmits individual data streams to multiple single-antenna users via an FRIS. In this case, the joint optimization of the positions and phase shifts of the FRIS element, as well as the BS precoding to maximize the sum-rate is studied. To solve the problem, a combination of techniques including PSO, semi-definite relaxation (SDR), and minimum mean square error (MMSE) is proposed. Numerical results demonstrate that the proposed FRIS approach significantly outperforms conventional RIS configurations in terms of achievable rate performance. 
\end{abstract}

\begin{IEEEkeywords}
Fluid antenna system (FAS), movable antenna, particle swarm optimization, programmable metasurface, reconfigurable antenna, reconfigurable intelligent surface (RIS).
\end{IEEEkeywords}

\section{Introduction}
\IEEEPARstart{I}{n} recent years, the technology of reconfigurable intelligent surfaces (RISs) has become the center of research in wireless communications. This is because it promises to enhance coverage of small base stations (BSs) in the sixth generation (6G) of wireless networks when the direct links can be severely weakened or blocked, a situation that can often happen when increasing the carrier frequency~\cite{Ref1new,RefEE,HMIMO-2024}. An RIS is a flat surface consisting of a large number of metamaterials with dynamically tunable responses over the impinging signals. By controlling the phase responses of the elements, RISs can intelligently reflect signals and hence have the ability to engineer the propagation environment that benefits many wireless applications~\cite{RIS-SRE-2023}. RISs are supposed to be of low power consumption comprising almost passive tunably reflecting elements, though recent trend also investigates the use of semi-passive~\cite{R-RIS-2020, HRIS-2024} and active~\cite{Long-ARIS-2021,Amplifying-RIS-2022,Rao-2023} RISs.

Despite the prospect of RIS, practical applications do face certain challenges. To unleash the power of RISs, the complexity of channel estimation must be overcome and is challenging as it involves the incoming and outgoing channels \cite{Wei-2021,Jian-2022,Jamali-2022}, and the computational power of an RIS is supposed to be low. Besides, the phase shifting resolution of the elements is often limited to keep the cost down~\cite{RIS-SRE-2023}, which complicates the optimization and reduces the degrees-of-freedom (DoF) \cite{Di-2020,Mesec}. The use of RISs in multi-band situations is also messy since it has dissimilar responses at different frequencies \cite{Jiang-2022}. Recently, \cite{MEH-2024, Wang-2024} examined the power consumption of RISs with practical models, with the latter work providing an experimental validation. Overall, a lot of issues around RISs have to do with whether the performance gain justifies the cost. It would be appealing if we could obtain higher DoF without the need for raising the number of RIS elements, since this scaling links to the associated cost, control overhead~\cite{RIS-control}, and overall complexity.

In terms of DoF, recent interest has witnessed efforts taking advantage of a new form of reconfigurable antennas, referred to as a fluid antenna system (FAS), to enhance the DoF in the physical layer of wireless communications~\cite{New2024aTutorial,Lu-2025}. In short, FAS represents the antenna technology of shape and position reconfigurability which is motivated by flexible structures, such as liquid-based antennas \cite{huang2021liquid,shen2024design} and liquid-metal metasurfaces~\cite{LM-RIS}, movable arrays \cite{basbug2017design,zhu2024historical}, metamaterial-based antennas \cite{johnson2015sidelobe,hoang2021computational,DMA-2023}, pixel reconfigurable antennas \cite{zhang2024pixel,Shen-2024}, etc. The latter two designs are particularly useful as their reconfigurability comes with no delay.

The concept of FAS was first introduced by Wong {\em et al.}~in 2020 \cite{Wong-2020fas,Wong-2021fas} which hypothesized a wireless communication system with a position-flexible FAS receiver. Since then, much effort has been made to understand the various performance limits of FAS channels, e.g., \cite{Khammassi2023,Vega2023novel,Vega2023asimple,Alvim2023on,Psomas2023continuous,New2023fluid}. A recent study has also investigated the diversity-multiplexing tradeoff of a multiple-input multiple-output (MIMO) channel with fluid antennas at both ends \cite{new2023information}. Evidently, channel state information (CSI) is essential to enable FAS and several CSI estimation schemes for FAS have recently been proposed \cite{xu2024channel,New-2025ce,Xu-2025ce,zhang2023successive}. Position optimization (i.e., port selection) \cite{Chai-2022} and jointly with beamforming have also become an important technology for many applications, e.g., \cite{Zhang-2024wpt,Wang-2024isac,zhou2024fasisac,Zou-2024}.

An interesting development also sees FAS being employed for multiple access, referred to as fluid antenna multiple access (FAMA). The idea is that interference can be mitigated at the receiver side by activating the position where the interference signal suffers from a deep fade due to multipath propagation. Following this idea, different forms of FAMA techniques have been proposed, including fast \cite{Wong-2022fama,Wong-2023fama}, slow switching \cite{Wong-2023sfama} and with coding \cite{Hong-2025}. Port selection with some analog signal combining has also been considered to improve interference immunity \cite{Wong-2024cuma,Wong-2024cuma2}. Opportunistic scheduling and FAMA also can work together to support a good number of users without precoding at the BS nor interference cancellation \cite{Waqar-2024}.

Inspired by the position reconfigurability of FAS, we propose in this paper to apply this concept to RISs. Our proposal is also motivated by the fact that an RIS is  usually not space-limited, but the number of its elements has to be limited due to complexity reasons. One lesson FAS has taught the wireless communications community is that spatial diversity originates from the space dimension and can be obtained by a {\em single} `fluid' element.\footnote{By  `fluid' element, we mean that each RIS element has position reconfigurability in addition to phase control for the reflected signal. In a broader sense, however, a fluid element can also exhibit other types of reconfigurability, such as shape and size, as defined within the FAS framework. Thus, our FRIS is more general than RISs based on liquid metals, where shape modification is used to achieve different reflection states~\cite{Jack-LM-2024}.} Motivated by this FAS property, this paper proposes a new paradigm, referred to as fluid RIS (FRIS), in which the radiating surface area is divided by non-overlapping subareas, each of which is equipped with a fluid element that can freely appear in any point of the designated subarea, and can reflect the incident signal with an optimized phase shift. This architecture can fully utilize the permissible area without having to install more elements accordingly.

To shed light on the potential of FRIS, we first investigate the single-user single-input single-output (SU-SISO) scenario where a transmitter with a single fixed-position antenna (FPA) communicates to a receiver with an FPA via an FRIS. Assuming that the direct link is broken, we find the optimal positions of the FRIS elements by an approach based on particle swarm optimization. Then, we extend our study to FRIS-aided multi-user multiple-input single-output (MU-MISO) systems in which a multi-FPA BS transmits to multiple users each equipped with an FPA via an FRIS. In this case, we solve the joint optimization problem of the FRIS element positions, their phase shifts, and the BS precoding, assuming the absence of a direct link. It is worth pointing out that FRIS was first proposed in \cite{Ye-2025} where integrated sensing and communication (ISAC) was considered. The limitation of \cite{Ye-2025} is however that the work was based on line-of-sight (LoS) only channels, which is not diffidently general. Our main contributions are summarized as follows:
\begin{itemize}
\item We obtain an optimized FRIS design for SU-SISO communications utilizing a PSO-based approach to determine the optimal positions of FRIS elements. In our system model, each fluid element is restricted to one subarea to strike a balance between DoF and practicality.
\item Additionally, we propose a comprehensive framework for the joint optimization of the FRIS element positions, their phase shifts, and the BS precoding to maximize the sum-rate performance of MU-MISO systems. Specifically, a grid search and a PSO-based approach are developed to find the optimal positions of the FRIS elements in each subarea. Then, for the selected positions, the FRIS phase shifts are obtained by using the semi-definite relaxation (SDR) technique. After that, the minimum mean square error (MMSE) scheme is employed to design the precoder at the BS side.

\item We evaluate the impact of various system parameters, such as the transmit signal-to-noise ratio (SNR), and the number of FRIS elements on the achievable rates. Our results demonstrate the significant performance improvements of FRIS-aided communication systems in both SU-SISO and MU-MISO scenarios compared to that using conventional RIS configurations.
\end{itemize}

The remainder of the paper is organized as follows: Section~\ref{sec:System-Model} introduces the system model of SU-SISO and MU-MISO with the aid of FRIS. The problem formulation and proposed algorithms are given in Section \ref{sec:main}. Then Section~\ref{sec:results} provides the numerical results and evaluates the proposed algorithms. Finally, we conclude the paper in Section~\ref{sec:conclude}.

\section{System Model\label{sec:System-Model}}
\begin{figure*}[]
\centering
\includegraphics[width=0.8\linewidth]{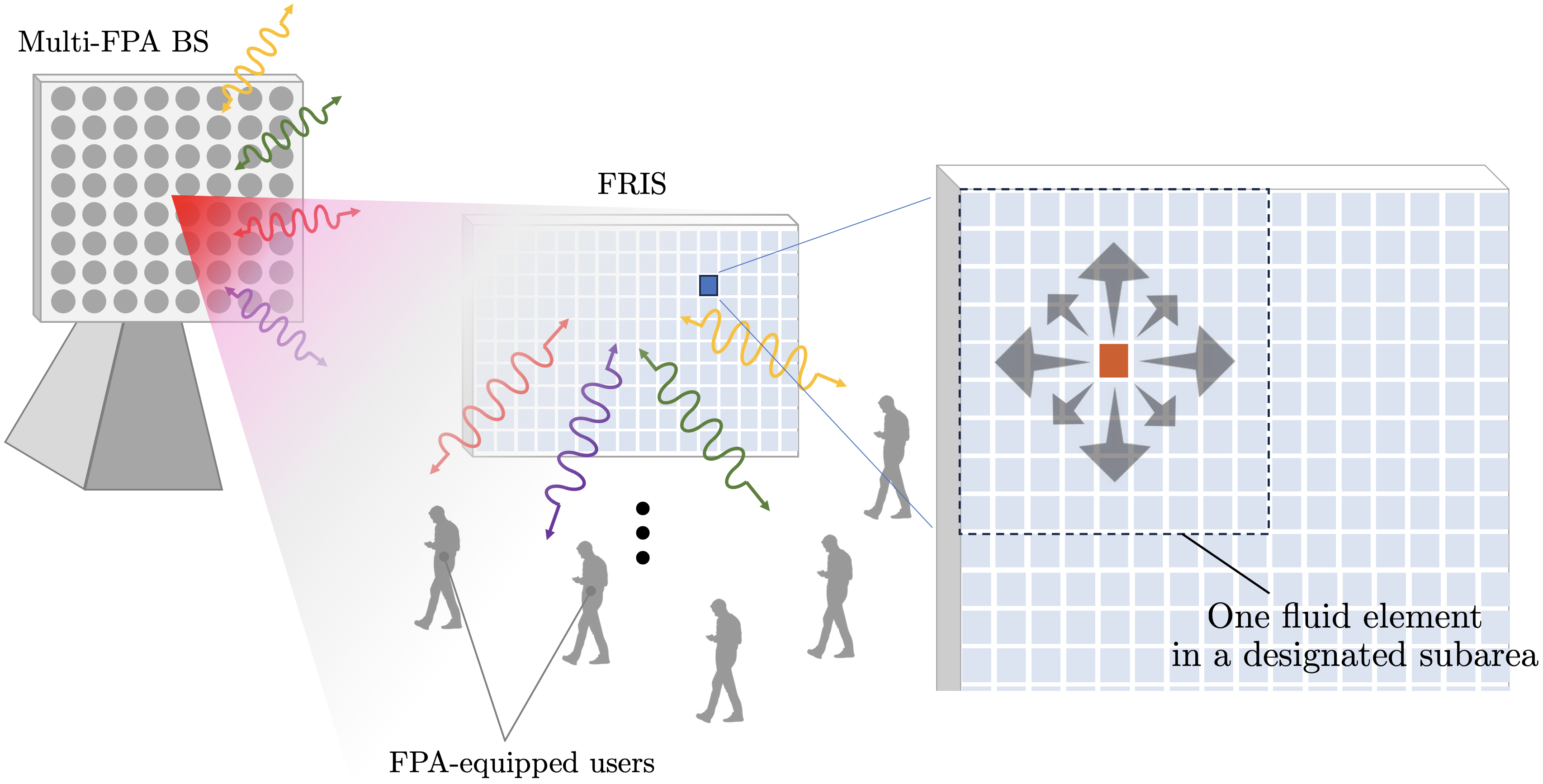}
\caption{Illustration of the MU-MISO system assisted by an FRIS.}\label{fig:fris_model}
\end{figure*}

In this paper, we consider two system models namely SU-SISO and MU-MISO. Both involve a transmitter needing a RIS to reestablish the communication channel to the receiver(s). In SU-SISO, the transmitter has a single FPA and there is only one user as receiver, while for the MU-MISO model depicted in Fig.~\ref{fig:fris_model}, the BS transmitter is equipped with multiple FPAs, aiming to communicate to many single-FPA users. For both cases, the direct link between the sender and receiver(s) is assumed broken and the use of RIS is necessary.

For the FRIS, its entire radiating area (${\rm A}\times{\rm A}$ in ${\rm m}^2$) is divided into $N$ non-overlapping subareas with no gaps. Each subarea is occupied by a fluid element, so the FRIS has totally $N$ fluid elements.\footnote{Although the FRIS is planar and two-dimensional, we will use an one-dimensional numbering for the elements to simplify our notation.} We use ${\cal S}_n$ to denote the set of all $(x,y)$-coordinates within the subarea of the $n$-th fluid element. The $n$-th fluid element is capable of switching to one of $L$ preset positions within ${\cal S}_n$ and we denote the position vector of the $n$-th fluid element by $\mathbf{t}_{n}=\left(x_{n},y_{n}\right)^{T}$ in which $T$ is the transpose operator. In other words, ${\bf t}_n$ dictates the point of reflection of the incident signal in ${\cal S}_n$. As in a typical RIS, the $n$-th fluid element will impose a phase shift $\angle\theta_{n}$ on the reflected signal. For convenience, we denote the positions of all the FRIS elements by $\mathbf{t}=\left[\mathbf{t}_{1,},\dots,\mathbf{t}_{N}\right]$ and their phase shifts by $\boldsymbol{\varTheta}={\rm diag}\left(\boldsymbol{\theta}\right)$ with $\boldsymbol{\theta}=\left[\theta_{1},\dots,\theta_{N}\right]^{T}$.

The model of our fluid element could be interpreted as an ideal movable point reflector, without considering the actual hardware, but in practice, this is more suitably realized using metamaterial-based and reconfigurable pixel technologies for their fast responses and much lower power consumption. For example, the slot-based technology in \cite{Liu-2025arxiv} can be used.

\subsection{SU-SISO\label{ssec:susiso}}
For the case of FRIS-assisted SU-SISO, the received signal at the user can be written as 
\begin{equation}\label{eq:siso}
y=\sqrt{P}\mathbf{g}^H\boldsymbol{\varTheta}\mathbf{h}s+\eta,
\end{equation}
where $\mathbf{h}\in\mathbb{C}^{N\times 1}$ is the channel between the transmitter and FRIS, $\mathbf{g}\in\mathbb{C}^{N\times1}$ is the channel between the FRIS and user, $s$ denotes the information-bearing signal with unit variance, $P$ is the transmit power of the BS, $\eta$ denotes the additive white Gaussian noise (AWGN) at the user, i.e., $\eta\sim {\cal CN}\left(0,\sigma^{2}\right)$, and the superscript $H$ is the Hermitian operator. For the phase shifting matrix $\boldsymbol{\varTheta}\in\mathbb{C}^{N\times N}$, we have $\theta_{n}=e^{j\phi_{n}}$, where $\phi_{n}\in[0,2\pi)$ is the phase shift of element $n$, and it is assumed that $\left|\theta_{n}\right|=1~\forall n$. As a result, the received SNR is given by
\begin{equation}
\gamma=\frac{P\left|\mathbf{g}^H\boldsymbol{\varTheta}\mathbf{h}\right|^{2}}{\sigma^{2}}.
\end{equation}

\subsection{MU-MISO\label{ssec:mumiso}}
If the BS is equipped with multiple FPAs and communicates to multiple users in the downlink, then the SU-SISO model becomes the MU-MISO model. Now, assuming that there are $M$ FPAs at the BS and $K$ users, the received signal at the $k$-th user can be expressed as
\begin{align}
y_{k} & = \mathbf{g}_{k}^{H}\boldsymbol{\varTheta}\mathbf{H}\mathbf{W}\mathbf{s}+\eta_{k},\notag\\
 & = \mathbf{g}_{k}^{H}\boldsymbol{\varTheta}\mathbf{H}\sum_{\ell=1}^K\mathbf{w}_{\ell}s_{\ell}+\eta_{k},
\end{align}
where ${\bf s}=[s_1\cdots s_K]^T$ with $s_\ell$ being the information signal for user $\ell$, $\eta_k$ denotes the AWGN at user $k$ defined similarly as before, ${\bf g}_k\in\mathbb{C}^{N\times 1}$ is the channel between the FRIS and user $k$, $\mathbf{H}\in\mathbb{C}^{N\times M}$ denotes the channel matrix from the BS and FRIS, and ${\bf w}_\ell\in\mathbb{C}^{M\times 1}$ is the BS precoding for user $\ell$. We also define ${\bf W}\triangleq [{\bf w}_1~{\bf w}_2\cdots {\bf w}_K]\in\mathbb{C}^{M\times K}$ as the overall precoding matrix of the BS. As such, the received signal-to-interference plus noise ratio (SINR) at user $k$ is found as
\begin{equation}
\gamma_{k}=\frac{P\left|\mathbf{g}_{k}^{H}\boldsymbol{\varTheta}\mathbf{H}\mathbf{w}_{k}\right|^{2}}{P\sum_{\ell=1\atop\ell\ne k}^K\left|\mathbf{g}_{k}^{H}\boldsymbol{\varTheta}\mathbf{H}\mathbf{w}_{\ell}\right|^{2}+\sigma^{2}}.
\end{equation}
For convenience, we also define ${\bf G}\triangleq[{\bf g}_1\cdots{\bf g}_K]$.

\subsection{Channel Model}\label{ssec:cmodel}
Here, we discuss how the channels are modelled. We focus on the channel matrices, ${\bf H}$ and ${\bf G}$, in the MU-MISO case since ${\bf h}$ and ${\bf g}$ in the SU-SISO case can be easily deduced from them. In the application of RIS, it is appropriate to model the channel between the BS and FRIS and that between the FRIS and users by Rician fading channels. Specifically, we have
\begin{align}
\mathbf{H}&=\sqrt{l_{b}}\left(\sqrt{\frac{\rho_{b}}{\rho_{b}+1}}\bar{\bf H}+\sqrt{\frac{1}{\rho_{b}+1}}\tilde{\bf H}\right),\\
\mathbf{g}_{k}&=\sqrt{l_{k}}\left(\sqrt{\frac{\rho_{k}}{\rho_{k}+1}}\bar{\bf g}_{k}+\sqrt{\frac{1}{\rho_{k}+1}}\tilde{\bf g}_{k}\right),
\end{align}
where $l_{b}$ and $l_{k}$ represent the path loss scaling factor between the BS and FRIS and that between the FRIS and user $k$, $\rho_{b}$ and $\rho_{k}$ are the Rician factors, $\bar{\bf H}$ and $\bar{\bf g}_{k}$ are the LoS components while $\tilde{\bf H}$ and $\tilde{\bf g}_{k}$ are the non-LoS (NLoS) components.

To characterize how the channel matrices change according to the positions of the elements at the FRIS, ${\bf t}$, we first consider the LoS components of the channels which are modelled as a function of the azimuth and elevation angle-of-departure (AoD) and angle-of-arrival (AoA) as
\begin{equation}
\bar{\bf H}=\mathbf{a}_{r}\left(\phi_{r}^{a},\phi_{r}^{e},\mathbf{t}\right)\mathbf{a}_{t}^{H}\left(\phi_{t}\right),
\end{equation}
where $\phi_t$ is the AoD from the BS and ${\bf a}_t(\cdot)$ is the BS transmit steering vector with
\begin{equation}
\left[\mathbf{a}_{t}\left(\phi_{t}\right)\right]_{m}=e^{j(m-1)\pi \sin\phi_{t}},
\end{equation}
where $[\cdot]_m$ returns the $m$-th entry of the input vector. Similarly, $\phi_{r}^{a}$ and $\phi_{r}^{e}$ represent, respectively, the azimuth AoA and the elevation AoA at the FRIS, and ${\bf a}_r(\cdot)$ is the receive steering vector at the FRIS with
\begin{equation}
\left[\mathbf{a}_{r}\left(\phi_{r}^{a},\phi_{r}^{e},\mathbf{t}\right)\right]_{n}=e^{j\frac{2\pi}{\lambda}\left(x_{n}\sin\phi_{r}^{a}\cos\phi_{r}^{e}+y_{n}\sin\phi_{r}^{e}\right)}
\end{equation}
with $\lambda$ being the carrier wavelength. For the FRIS to user $k$'s channel, we also have
\begin{equation}
\bar{\bf g}_{k}=\mathbf{a}\left(\phi_{k}^{a},\phi_{k}^{e},\mathbf{t}\right),
\end{equation}
where $\phi_{k}^{a}$ and $\phi_{k}^{e}$ are, respectively, the azimuth and elevation AoD from the FRIS to user $k$, and ${\bf a}(\cdot)$ is the transmit steering vector of the FRIS to user $k$ with 
\begin{equation}
\left[\mathbf{a}\left(\phi_{k}^{a},\phi_{k}^{e},\mathbf{t}\right)\right]_{n}=e^{j\frac{2\pi}{\lambda}\left(x_{n}\sin\phi_{k}^{a}\cos\phi_{k}^{e}+y_{n}\sin\phi_{k}^{e}\right)}.
\end{equation}

For the NLoS components, we model the spatial correlation of the channels using Jake's model. Let us define $\mathbf{J}_b$ and $\mathbf{J}_k$ as the spatial correlation matrices for the BS-FRIS and FRIS-user $k$ channels, respectively. Then we have
\begin{align}
[\mathbf{J}_b]_{i,j} &= J_0\left(\frac{2\pi \Delta d_{i,j}}{\lambda}\right),\\
[\mathbf{J}_k]_{m,n} &= J_0\left(\frac{2\pi \Delta d_{m,n}}{\lambda}\right)
\end{align}
where $J_0(\cdot)$ denotes the zeroth-order Bessel function of the first kind, and $\Delta d_{i,j}$ and $\Delta d_{m,n}$ denote the distances between the $i$-th and $j$-th FRIS elements of the BS-FRIS channel, and between the $m$-th and $n$-th FRIS elements of the FRIS-user $k$ channel, respectively. Specifically, the spatial correlation matrices $\mathbf{J}_b$ and $\mathbf{J}_k$ can be decomposed as
\begin{align}
\mathbf{J}_b &= \mathbf{U}_b \mathbf{\Lambda}_b \mathbf{U}_b^H,\\
\mathbf{J}_k &= \mathbf{U}_k \mathbf{\Lambda}_k \mathbf{U}_k^H,
\end{align}
where $\mathbf{U}_b$ and $\mathbf{U}_k$ are unitary matrices containing the eigenvectors of the correlation matrices, and $\mathbf{\Lambda}_b$ and $\mathbf{\Lambda}_k$ are diagonal matrices containing the eigenvalues. Thus, the mathematical expressions of the NLoS channel matrix $\tilde{\bf H}$ and the channel vector $\tilde{\bf g}_k$ can be expressed as
\begin{align}
\tilde{\bf H} &= \mathbf{\Lambda}_b^{1/2} \mathbf{U}_b^H \hat{\bf H},\\
\tilde{\bf g}_{k} &=  \mathbf{\Lambda}_{k}^{1/2} \mathbf{U}_{k}^H \hat{\bf g}_{k},
\end{align}
where $\hat{\bf H}$ and $\hat{\bf g}_k$ represent the uncorrelated small-scale fading components, with their elements modelled as independent and identically distributed (i.i.d.) complex Gaussian distributions with zero mean and unit variance.

\section{Main Results}\label{sec:main}
In this section, we present the problem formulations of both models and attempt to optimize the FRIS for performance enhancement. Specifically, our interest is on maximizing the achievable rate of the system given some constraints.

\subsection{SU-SISO}\label{ssec:siso}
First, the SU-SISO model is considered. Assuming capacity-achieving codec is being adopted, the achievable rate depends only on the SNR. Our objective is to maximize the achievable rate by optimizing the FRIS elements. That is,
\begin{subequations}\label{eq:siso-problem}
\begin{align}
\max_{\mathbf{t},\boldsymbol{\varTheta}} &~~\log_2\left(1+\frac{P\left|\mathbf{g}^H\boldsymbol{\varTheta}\mathbf{h}\right|^{2}}{\sigma^{2}}\right)\\
{\rm s.t.} &~~\mathbf{t}_{n}\in\mathcal{S}_{n},\forall n\label{eq:subarea}\\
&~~\left\Vert \mathbf{t}_{n}-\mathbf{t}_{n'}\right\Vert _{2}\geq D,\forall n\neq n'\label{eq:min-d}\\
&~~\left|\theta_{n}\right|=1,\forall n,
\end{align}
\end{subequations}
in which $\mathcal{S}_{n}$ is the feasible subarea for the $n$-th fluid element to roam and $D$ denotes a minimum spacing set between two fluid elements. Depending on what technology the fluid element is based on, the value of $D$ could have a different interpretation. For example, if mechanically movable radiating elements are employed, then $D$ will play a key role to control the level of mutual coupling between the adjacent elements.\footnote{Note that mutual coupling is not always harmful as observed in \cite{Wong-2024cuma2} as far as beamforming is concerned. Even if the aim is to minimize mutual coupling, state-of-the-art technologies can already reduce it to a very low level even if the separation is as little as $0.1\lambda$ where $\lambda$ is the carrier wavelength \cite{Murch-2007,Althuwayb-2023}.} By contrast, $D$ would be much less meaningful if other technologies such as metamaterials or reconfigurable pixels are considered.

To solve (\ref{eq:siso-problem}), it is clear that the optimal phase shift of each fluid element is to cancel the overall phase shift imposed by the first- and second-hop channels, i.e., 
\begin{equation}
\theta_{n}^{*}=e^{-j\left(\angle\left[\mathbf{h}\right]_{n}+\angle\left[\mathbf{g}\right]_{n}\right)}.
\end{equation}
Hence, the optimization problem (\ref{eq:siso-problem}) can be reduced to
\begin{subequations}\label{eq:siso-problem2}
\begin{align}
\max_{\mathbf{t}} &~~\log_2\left(1+\frac{P\left|\sum_{n=1}^N\left|[{\bf h}_n]\right|\left|[{\bf g}_n]\right|\right|^{2}}{\sigma^{2}}\right)\label{eq:rate-siso}\\
{\rm s.t.} &~~\mathbf{t}_{n}\in\mathcal{S}_{n},\forall n\label{eq:subarea-siso}\\
&~~\left\Vert \mathbf{t}_{n}-\mathbf{t}_{n'}\right\Vert _{2}\geq D,\forall n\neq n'.\label{eq:min-d-siso}
\end{align}
\end{subequations}

The problem is inherently non-convex and computationally intensive due to the high-dimensional search space. Evaluating all the possible combinations of the fluid element positions by adopting an exhaustive search algorithm or greedy algorithm is prohibitively complex, especially when the number of fluid elements grows large. To efficiently overcome this challenge, a PSO-based algorithm is proposed. PSO can converge quickly to near-optimal solutions without requiring derivative information. Moreover, it can incorporate subarea boundaries and minimum spacing between elements into the optimization, ensuring feasible and deployable solutions. PSO is an efficient choice for optimizing FRIS due to its efficiency in handling high-dimensional, non-convex, and combinatorial problems. Unlike exhaustive search, which is computationally prohibitive, PSO leverages swarm intelligence to explore the solution space effectively while avoiding local optima.  Compared to gradient-based methods, which struggle with non-differentiable constraints, and alternative metaheuristics like genetic algorithms and simulated annealing, which often exhibit slower convergence, PSO provides a balance between global exploration and local refinement. Its ability to incorporate spatial constraints ensures practical deployment, while its fast convergence and computational efficiency enable near-optimal solutions in real-time wireless scenarios. 

In our model, the area of FRIS is divided into $N$ subareas, each assigned to one fluid element. Each subarea is defined by some boundary. For the $n$-th subarea, it is defined by the ranges $[x_{\min}^{n},x_{\max}^{n}]$ and $[y_{\min}^{n},y_{\max}^{n}]$ in the $x$- and $y$-axis, respectively. The distance between any two subareas should satisfy the constraint (\ref{eq:min-d-siso}). The optimization within each subarea is conducted independently using PSO. In the PSO-based algorithm, we first randomly initialize particles with the positions and velocities. Then we calculate the fitness of each particle based on the objective function. Each particle updates its position based on the individual experience (the currently known local best-position, $\mathbf{t}_{i,{\rm best}}$) and the swarm experience (the currently known global best position, $\mathbf{t}_{\rm best}$). Consequently, the velocity of each particle is updated by
\begin{equation}
\mathbf{v}_{i}^{\left(t+1\right)}=\omega\mathbf{v}_{i}^{\left(t\right)}+c_{1}r_{1}\left[\mathbf{t}_{i,{\rm best}}-\mathbf{t}_{i}^{\left(t\right)}\right]+c_{2} r_{2}\left[\mathbf{t}_{\rm best}-\mathbf{t}_{i}^{\left(t\right)}\right],
\end{equation}
where $t$ is the iteration index, $\mathbf{v}_{i}^{\left(t+1\right)}$ is the new velocity of particle $i$ at iteration $t+1$, $\omega$ is the inertia weight to balance the speed and accuracy of particle swarm search, $c_{1},c_{2}$ are the individual and global learning factors, which represent the step size of each particle moving toward the best position, $r_{1},r_{2}$ are random numbers between $0$ and $1$ to enhance the randomness of the search for escaping from local optima. Also, for each iteration, the position of each particle is updated as
\begin{equation}
\mathbf{t}_{i}^{\left(t+1\right)}={\cal P}\left(\mathbf{t}_{i}^{\left(t\right)}+\mathbf{v}_{i}^{\left(t+1\right)}\right),
\end{equation}
where ${\cal P}\left(\cdot\right)$ is the projection function to enforce the boundary constraint (\ref{eq:subarea-siso}). If a particle moves out of the boundaries of the feasible region, we project its position to the corresponding maximum or minimum value. Also, constraint (\ref{eq:min-d-siso}) is already satisfied after dividing the FRIS area into $N$ disjoint subareas with minimum spacing $D$. Moreover, we introduce an adaptive penalty factor to the fitness function to guarantee that the constraints are met. Thus, the objective function becomes
\begin{equation}
\mathfrak{O}(\mathbf{t}_{i}^{\left(t\right)})=R(\mathbf{t}_{i}^{\left(t\right)})-\tau\mathfrak{B}(\mathbf{t}_{i}^{\left(t\right)}),
\end{equation}
in which $R(\cdot)$ represents the original rate function in (\ref{eq:rate-siso}), $\tau$ is a large positive penalty parameter which ensures that the inequality, $R(\mathbf{t}_{i}^{\left(t\right)})-\tau\leq0$ holds for all the positions, meaning that the penalty parameter moves each particle position in order to ensure the minimum spacing constraint, and $\mathfrak{B}(\mathbf{t}_{i}^{\left(t\right)})$ is the penalty function which accounts for the position violation of fluid element $i$ against (\ref{eq:min-d-siso}), defined as
\begin{equation}
\mathfrak{B}\left(\mathbf{t}\right)=\left\{\left(\mathbf{t}_{n},\mathbf{t}_{n'}\right)\left|\left|\mathbf{t}_{n}-\mathbf{t}_{n'}\right\Vert _{2}<D, 1\le n<n'\le N\right.\right\}.
\end{equation}
The PSO-based algorithm is summarized in Algorithm \ref{alg:PSO-siso}.

\begin{algorithm}[]
1. Input: Overall FRIS area $[X_{\min},X_{\max}]$ and $[Y_{\min},Y_{\max}]$, $N$, $D$, PSO parameters.

2. Divide the FRIS area into $N$ subareas with distance $D$, each defined by $[x_{\min}^{n},x_{\max}^{n}]$ and $[y_{\min}^{n},y_{\max}^{n}]$.

3. Pre-compute the channels for each subarea. 

4. For each FRIS element: 

$\:$$\:$$\:$a. Perform PSO within the subarea: 

$\:$$\:$$\:$$\:$$\:$$\:$i. Initialize particles and velocities. 

$\:$$\:$$\:$$\:$$\:$$\:$ii. For each iteration: 

$\:$$\:$$\:$$\:$$\:$$\:$$\:$$\:$$\:$-- Evaluate fitness for all particles. 

$\:$$\:$$\:$$\:$$\:$$\:$$\:$$\:$$\:$-- Update personal and global best positions. 

$\:$$\:$$\:$$\:$$\:$$\:$$\:$$\:$$\:$-- Update particle velocities and positions. 

$\:$$\:$$\:$$\:$$\:$$\:$$\:$$\:$$\:$-- Enforce subarea boundaries. 

$\:$$\:$$\:$$\:$$\:$$\:$iii. Select the optimal position for the element. 

$\:$$\:$$\:$b. Check minimum spacing constraint (\ref{eq:min-d-siso}) with previously selected positions. 

$\:$$\:$$\:$c. If spacing constraint is violated: 

$\:$$\:$$\:$$\:$$\:$$\:$$\:$$\:$$\:$-- Re-optimize for a valid position. 

$\:$$\:$$\:$d. Append valid position to selected positions. 

5. Output: selected positions and corresponding rate values.

\protect\caption{PSO-Based Algorithm}\label{alg:PSO-siso}
\end{algorithm}

\subsection{MU-MISO}
In the case of FRIS-aided MU-MISO system, our interest is to maximize the sum-rate of all users in the downlink, which can be formulated as
\begin{subequations}\label{eq:miso-problem}
\begin{align}
\max_{\mathbf{t},\boldsymbol{\varTheta},{\bf W}} &~~\sum_{k=1}^K\log_2\left(1+\frac{\left|\mathbf{g}_{k}^{H}\boldsymbol{\varTheta}\mathbf{H}\mathbf{w}_{k}\right|^{2}}{\sum_{\ell=1\atop \ell\ne k}^K\left|\mathbf{g}_{k}^{H}\boldsymbol{\varTheta}\mathbf{H}\mathbf{w}_{\ell}\right|^{2}+\sigma^{2}}\right)\\
{\rm s.t.} &~~\mathbf{t}_{n}\in\mathcal{S}_{n},\forall n\\
&~~\left\Vert \mathbf{t}_{n}-\mathbf{t}_{n'}\right\Vert _{2}\geq D,\forall n\neq n'\\
&~~\left|\theta_{n}\right|=1,\forall n,\\
&~~\sum_{k=1}^K\|{\bf w}_k\|^2\le P,\label{eq:pcon}
\end{align}
\end{subequations}
where (\ref{eq:pcon}) is the power constraint which becomes important in the optimization. The other constraints remain the same. An obvious challenge now is to also jointly optimize the precoding matrix at the BS. Problem (\ref{eq:miso-problem}) is very challenging, due to the complexity of the objective function, the modulus constraint, and the non-convex norm form. Given the non-convexity of the problem, we resort to an alternating optimization strategy. In particular, we solve (\ref{eq:miso-problem}) by updating each variable iteratively with the other variables fixed. Now we decouple it into three sub-problems: 1) FRIS Position Optimization, 2) Phase Shift Optimization, and 3) Precoding Optimization. 

\begin{enumerate}
\item {\em FRIS Position Optimization}--We aim to determine the optimal positions of the fluid elements that maximize the received SINR and improve the overall system performance. The phase shifts and precoder are assumed to be fixed or precomputed based on the current positions. For each fluid element, we consider a set of possible positions within a predefined subarea, and the position that maximizes the sum rate can be selected using PSO, as explained in Section \ref{ssec:siso}, or by a grid search which systematically evaluates each candidate position within a defined search space. During the grid search, candidate positions that violate the spacing constraint are excluded from the search. For fluid element $n$, divide its subarea into a grid with resolution $\Delta x$ and $\Delta y$ in both directions, producing a set of candidate positions. For candidate position $\mathbf{t}_{n}$, \LyXZeroWidthSpace compute the BS-to-FRIS channel $\mathbf{H}^{(n)}$ and the FRIS-to-users channel $\mathbf{G}^{(n)}$. Then compute the sum-rate for each candidate position. For each subarea, select the position $\mathbf{t}_{n}^{*}$ that maximizes the sum-rate, i.e.,
\begin{equation}\label{eq:sumrate-t}
\mathbf{t}_{n}^{*}=\arg\max_{\mathbf{t}_{n}}\sum_{k=1}^{K}\log_{2}\left(1+{\rm SINR}_{k}(\mathbf{t}_{n})\right).
\end{equation}

PSO is computationally efficient for large search spaces, when the grid search becomes expensive. The key components of PSO to find the optimal positions has been explained in Section \ref{ssec:siso}.

\item {\em Phase Shift Optimization}--Here we optimize the phase shift matrix $\boldsymbol{\varTheta}$ to align the reflected signals constructively. Thus, the optimization problem is formulated as
\begin{subequations}\label{eq:miso-problem-phase}
\begin{align}
\max_{\boldsymbol{\varTheta}} &~~\sum_{k=1}^K\log_2\left(1+\frac{\left|\mathbf{g}_{k}^{H}\boldsymbol{\varTheta}\mathbf{H}\mathbf{w}_{k}\right|^{2}}{\sum_{\ell=1\atop \ell\ne k}^K\left|\mathbf{g}_{k}^{H}\boldsymbol{\varTheta}\mathbf{H}\mathbf{w}_{\ell}\right|^{2}+\sigma^{2}}\right)\\
{\rm s.t.} &~~\left|\theta_{n}\right|=1,\forall n,
\end{align}
\end{subequations}
which can be solved using SDR techniques. SDR can efficiently handle the non-convex unit-modulus constraint on the FRIS elements, which makes direct optimization difficult. By lifting the problem into a higher-dimensional space and relaxing the rank-one constraint, SDR transforms it into a convex semi-definite program (SDP) that can be efficiently solved using standard optimization tools.  Compared to brute-force or non-linear methods, SDR is computationally efficient and enables iterative refinements for higher spectral efficiency and sum rate. Let us define the effective channel for user $k$, $\mathbf{h}_{{\rm eff},k}=\mathbf{g}_{k}^{H}\boldsymbol{\varTheta}\mathbf{H}$. Thus, the SINR for user $k$ is reexpressed as
\begin{equation}
\gamma_{k}=\frac{\left|\mathbf{h}_{{\rm eff},k}\mathbf{w}_{k}\right|^{2}}{\sum_{\ell=1\atop \ell\neq k}^K\left|\mathbf{h}_{{\rm eff},k}\mathbf{w}_{\ell}\right|^{2}+\sigma^{2}}.
\end{equation}

To relax the unit-modulus constraint, we write $\mathbf{v}=[e^{j\phi_{1}},\dots,e^{j\phi_{N}}]^{T}$, and $\boldsymbol{\varTheta}={\rm diag}(\mathbf{v})$, which gives
\begin{equation}
\mathbf{h}_{{\rm eff},k}=\mathbf{v}^{H}{\rm diag}\left(\mathbf{g}_{k}^{H}\right)\mathbf{H}
\end{equation}
and
\begin{equation}
\left|\mathbf{h}_{{\rm eff},k}\mathbf{w}_{k}\right|^{2}=\mathbf{v}^{H}\mathbf{A}_{k}\mathbf{v},
\end{equation}
where $\mathbf{A}_{k}={\rm diag}\left(\mathbf{g}_{k}^{H}\right)\mathbf{H}\mathbf{w}_{k}\mathbf{w}_{k}^{H}\mathbf{H}^{H}{\rm diag}\left(\mathbf{g}_{k}^{H}\right)^{H}$. Similarly, we get
\begin{equation}
\sum_{j\neq k}\left|\mathbf{h}_{{\rm eff},k}\mathbf{w}_{j}\right|^{2}=\sum_{j\neq k}\mathbf{v}^{H}\mathbf{B}_{k,j}\mathbf{v},
\end{equation}
in which $\mathbf{B}_{k,j}={\rm diag}\left(\mathbf{g}_{k}^{H}\right)\mathbf{H}\mathbf{w}_{j}\mathbf{w}_{j}^{H}\mathbf{H}^{H}{\rm diag}\left(\mathbf{g}_{k}^{H}\right)^{H}$. The overall SINR term then becomes
\begin{equation}
\gamma_{k}=\frac{\mathbf{v}^{H}\mathbf{A}_{k}\mathbf{v}}{\mathbf{v}^{H}\mathbf{B}_{k}\mathbf{v}+\sigma^{2}}.
\end{equation}
where $\mathbf{B}_{k}\triangleq\sum_{j\neq k}\mathbf{B}_{k,j}$. The next step is to replace $\mathbf{v}\mathbf{v}^{H}$ by a positive semidefinite (PSD) matrix $\mathbf{V}$, so $\mathbf{V}=\mathbf{v}\mathbf{v}^{H}$. SDR removes the rank-one constraint, allowing $\mathbf{V}$ to be any PSD matrix, and the solution is projected back to the feasible set. Now we can reformulate the phase shift optimization problem as 
\begin{subequations}\label{eq:miso-problem-phase2}
\begin{align}
\max_{\bf V} &~~\sum_{k=1}^K\log_2\left(1+\frac{{\rm trace}\left(\mathbf{A}_{k}\mathbf{V}\right)}{{\rm trace}\left(\mathbf{B}_{k}\mathbf{V}\right)+\sigma^{2}}\right)\\
{\rm s.t.} &~~\mathbf{V}\succeq\mathbf{0},\\
&~~{\rm diag}\left(\mathbf{V}\right)=1.
\end{align}
\end{subequations}

Then we can replace the fraction inside the logarithm using an alternative transformation. Hence, the objective function can be written as
\begin{equation}
\log_{2}\left({\rm trace}\left(\mathbf{A}_{k}\mathbf{V}\right)+{\rm trace}(\mathbf{B}_{k}\mathbf{V})+\sigma^{2}\right).
\end{equation}
The relaxed SDP problem can then be written as
\begin{subequations}\label{eq:miso-problem-phase3}
\begin{align}
\max_{\bf V} &~~\sum_{k=1}^K\log_{2}\left({\rm trace}\left(\mathbf{A}_{k}\mathbf{V}\right)+{\rm trace}(\mathbf{B}_{k}\mathbf{V})+\sigma^{2}\right)\label{eq:obj-phase}\\
{\rm s.t.} &~~\mathbf{V}\succeq\mathbf{0},\\
&~~{\rm diag}\left(\mathbf{V}\right)=1.
\end{align}
\end{subequations}
To proceed further, we approximate (\ref{eq:obj-phase}) using first-order Taylor series expansion. Before that, we define 
\begin{equation}
f_{k}\left(\mathbf{V}\right)\triangleq{\rm trace}\left(\mathbf{A}_{k}\mathbf{V}\right)+{\rm trace}(\mathbf{B}_{k}\mathbf{V})+\sigma^{2}.
\end{equation}
By approximating $\log_{2}\left(f\left(\mathbf{V}\right)\right)$ using first-order Taylor series expansion around a feasible point $\mathbf{V}^{\left(t-1\right)}$ (at the previous iteration), we have
\begin{multline}
\log_{2}\left(f_{k}\left(\mathbf{V}\right)\right)\approx\log_{2}\left(f_{k}(\mathbf{V}^{\left(t-1\right)})\right)+\\
\frac{1}{\ln2}\frac{{\rm trace}\left(\mathbf{A}_{k}\mathbf{V}\right)+{\rm trace}(\mathbf{B}_{k}\mathbf{V})-f_{k}\left(\mathbf{V}^{\left(t-1\right)}\right)}{f_{k}\left(\mathbf{V}^{\left(t-1\right)}\right)}.
\end{multline}

This transforms the problem into a convex one that can be solved iteratively. Generally, the relaxed problem may not lead to a rank-one solution, which implies that the optimal objective value of problem can only serve as an upper bound. Additional steps are needed to construct a rank-one solution from the obtained higher-rank solution to the problem. To do so, we extract a feasible solution from $\mathbf{V}$ using eigenvalue decomposition
\begin{equation}\label{eq:evd}
\left\{\begin{aligned}
\mathbf{v}&=\sqrt{\lambda_{1}}\mathbf{u}_{1},\\
\mathbf{V}&\approx\lambda_{1}\mathbf{u}_{1}\mathbf{u}_{1}^{H},
\end{aligned}\right.
\end{equation}
where $\lambda_{1}$ and $\mathbf{u}_{1}$ are the largest eigenvalue and the corresponding eigenvector of $\mathbf{V}$. The steps of the phase shift optimization are summarized in Algorithm \ref{alg:phase}.

\begin{algorithm}
1. Initialize $\mathbf{V}^{(0)}$ with a feasible PSD matrix. 

2. Repeat until convergence:

$\:$$\:$$\:$ Compute $f_{k}(\mathbf{V}^{(t-1)})$ for the current iteration. 

$\:$$\:$$\:$ Solve the convex SDP problem. 

$\:$$\:$$\:$ Update $\ensuremath{\mathbf{V}^{(t)}}$ with the solution from SDR. 

$\:$$\:$$\:$ If $\|\mathbf{V}^{(t)}-\mathbf{V}^{(t-1)}\|$ is small, stop.

3. Use eigenvalue decomposition (\ref{eq:evd}). 

\protect\caption{Phase Shift Optimization.}\label{alg:phase}
\end{algorithm}

\item {\em Precoding Matrix Optimization}--The remaining task is to optimize the precoding matrix $\mathbf{W}$ at the BS, which can be formulated as 
\begin{subequations}\label{eq:miso-problem-W}
\begin{align}
\max_{{\bf W}} &~~\sum_{k=1}^K\log_2\left(1+\frac{\left|\mathbf{g}_{k}^{H}\boldsymbol{\varTheta}\mathbf{H}\mathbf{w}_{k}\right|^{2}}{\sum_{\ell=1\atop \ell\ne k}^K\left|\mathbf{g}_{k}^{H}\boldsymbol{\varTheta}\mathbf{H}\mathbf{w}_{\ell}\right|^{2}+\sigma^{2}}\right)\\
{\rm s.t.} &~~\sum_{k=1}^K\|{\bf w}_k\|^2\le P.\label{eq:pcon-W}
\end{align}
\end{subequations}
To solve this precoding optimization problem, we implement the MMSE technique. MMSE minimizes the total error between the transmitted and received signals, leading to better robustness. It is particularly beneficial in multi-user systems, where interference between users should be mitigated while maximizing the sum rate. Additionally, MMSE is convex and computationally efficient, allowing for real implementation in FRIS-assisted MU-MISO networks. Inspired by the MU-MISO results in \cite{zhang2023successive}, we define the mean square error as
\begin{equation}
e_{k}=1-2{\rm Re}\left\{u_{k}^*\boldsymbol{h}_{{\rm eff},k}\mathbf{w}_{k}\right\}+|{u}_{k}|^2 Q_{k},
\end{equation}
where $u_{k}$ is the complex receiver scaling at user $k$, $Q_k=\sum_{\ell=1\atop \ell\ne k}^K\left|\mathbf{g}_{k}^{H}\boldsymbol{\varTheta}\mathbf{H}\mathbf{w}_{\ell}\right|^{2}+\sigma^{2}$ is the total interference-plus-noise power for user $k$, and
\begin{equation}
\boldsymbol{h}_{{\rm eff},k}\triangleq \mathbf{g}_{k}^{H}\boldsymbol{\varTheta}\mathbf{H}.
\end{equation}
Then by introducing the auxiliary optimization variables $\boldsymbol{\varpi}=\left[\varpi_{1},\dots,\varpi_{K}\right]^{T}$, we can reformulate (\ref{eq:miso-problem-W}) as a more tractable MMSE problem as 
\begin{subequations}\label{eq:W-mmse}
\begin{align}
\min_{{\bf W},\{u_k\},\boldsymbol{\varpi}} &~~\sum_{k=1}^K\left(\varpi_{k}e_{k}-\log\left(\varpi_{k}\right)\right)\\
{\rm s.t.} &~~\sum_{k=1}^K\|{\bf w}_k\|^2\le P\\
&~~\varpi_{k}>0,\forall k,
\end{align}
\end{subequations}
which can be solved using an iterative MMSE algorithm below. To do so, first, we optimize $u_{k}$ for a fixed $\mathbf{W}$ and $\boldsymbol{\varpi}$, with the optimal solution given by
\begin{equation}
u_{k}=\frac{\boldsymbol{h}_{{\rm eff},k}\mathbf{w}_{k}}{Q_k}.
\end{equation}
Then for fixed $\mathbf{W}$ and $\{u_k\}_{\forall k}$, the optimal auxiliary variable can be found as
\begin{equation}
\varpi_{k}=\frac{1}{e_{k}}.
\end{equation}
Now, for fixed $\{u_k\}_{\forall k}$ and $\boldsymbol{\varpi}$, the optimal precoder can be obtained by solving
\begin{subequations}\label{eq:W-mmse2}
\begin{align}
\min_{{\bf W}} &~~\sum_{k=1}^K\varpi_{k}\left[1-2{\rm Re}\left\{u_{k}^*\boldsymbol{h}_{{\rm eff},k}\mathbf{w}_{k}\right\}+|{u}_{k}|^2 Q_{k}\right]\\
{\rm s.t.} &~~\sum_{k=1}^K\|{\bf w}_k\|^2\le P,
\end{align}
\end{subequations}
which is convex in $\mathbf{W}$ and can therefore be solved using standard convex optimization tools (e.g., CVX). As such, we have Algorithm \ref{alg:W} to solve (\ref{eq:miso-problem-W}).

\begin{algorithm}[H]
1. Generate possible positions $\ensuremath{\mathbf{t}}$ for each fluid element.

2. Precompute BS-to-FRIS and FRIS-to-user channels $\mathbf{H}$ and $\ensuremath{\mathbf{G}}$ for all possible positions.

3. For each FRIS element, select the position using (\ref{eq:sumrate-t}).

4. Use Algorithm \ref{alg:phase} to obtain optimal phase shifts of the selected positions.

5. Use MMSE Algorithm to obtain $\mathbf{W}$.

6. Repeat Step 3 until convergence in the sum-rate.

\protect\caption{Proposed Algorithm for BS Precoding.}\label{alg:W}
\end{algorithm}
\end{enumerate}

\subsection{Complexity and Convergence Analysis}
The complexity of FRIS optimization with $N$ fluid elements and $\mathcal{N}$ candidate positions in each subarea is $O(N\cdot\mathcal{N})$. The complexity of phase shift optimization using SDR is $O(N^{3})$ per iteration, and the complexity of precoding optimization using MMSE is $O(M^{3}+M^{2}K)$ per iteration. Therefore, the overall complexity per iteration is $O(K\cdot N^{3}+M^{3})$.

The FRIS position optimization converges in a few steps due to grid search, and the phase shift optimization using SDR converges to a solution satisfying the relaxed constraints. In addition, the precoding optimization using MMSE converges due to the convexity of sub-problems. The overall objective (sum-rate) is non-decreasing and upper-bounded. 

\section{Numerical Results and Discussion}\label{sec:results}
In this section, simulation results are provided to assess the rate performance of the proposed FRIS-assisted systems and compare that with the conventional RIS counterpart. In the simulations, we consider the locations of the BS and FRIS at $(0,0)$ and $(20,20)$ in meters, respectively, while the users are distributed in a circle centered at $(20,0)$ meters with radius of $10~{\rm m}$. The minimal element spacing is $D=\lambda/2$, and the size of FRIS/RIS is ${\rm A}\times{\rm A}$ where ${\rm A}=4~{\rm m}$ unless otherwise specified. The wavelength of $\lambda=0.125~{\rm m}$ is set and thus the carrier frequency is $2.4~{\rm GHz}$. Moreover, for the case of MU-MISO, we have $M=8$ BS antennas and $K=4$ users. Additionally, the path-loss exponent is set to $2.7$, and the Rician factor is $3$. The AoA and AoD of the BS and FRIS are randomly and uniformly distributed in $(0,\pi)$.

\subsection{SU-SISO}
\begin{figure}[]
\begin{center}
\subfigure[Optimal elements positions]{\includegraphics[width=\columnwidth]{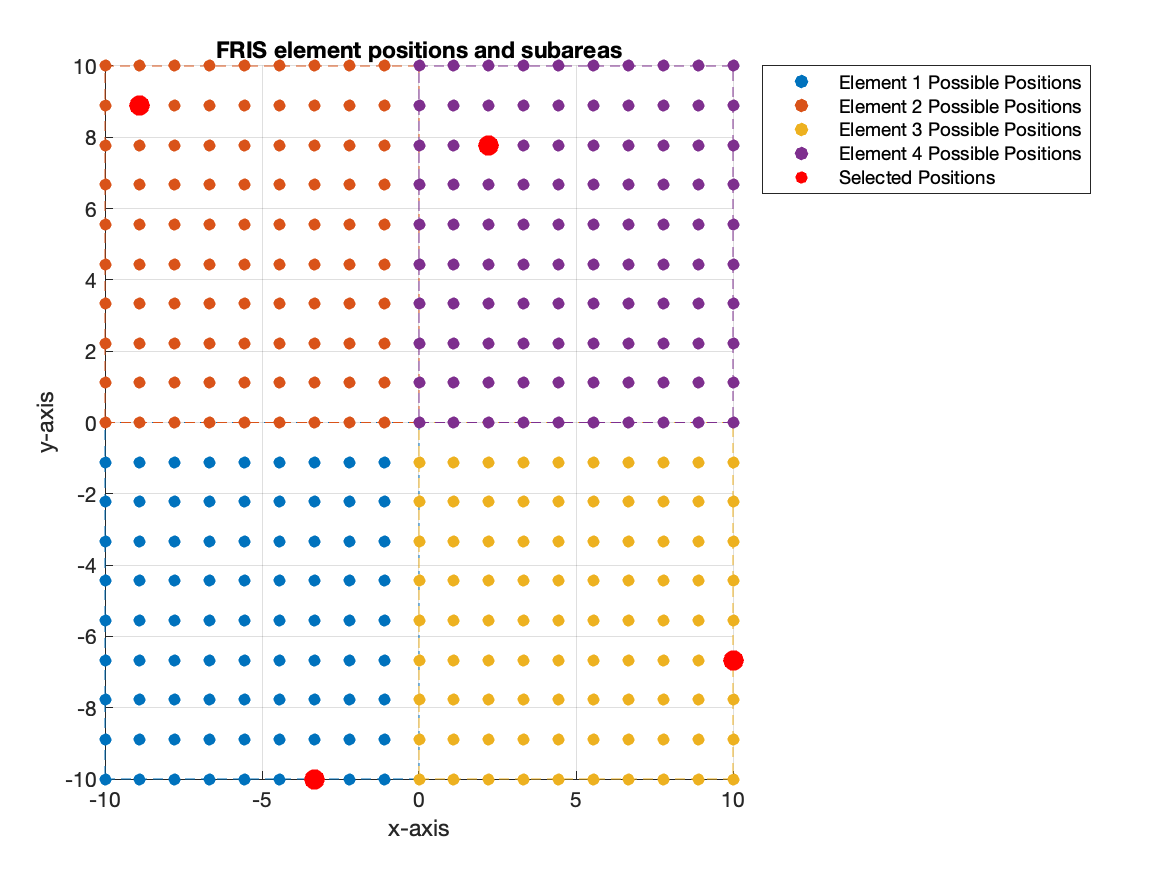}\label{fig2:N4pos}}
\subfigure[Rate performance]{\includegraphics[width=.95\columnwidth]{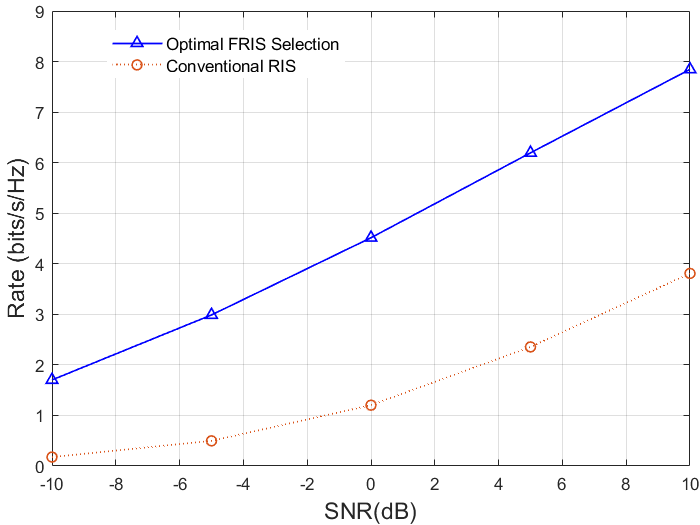}\label{fig2:N4rate}}
\caption{Comparison between the proposed FRIS-aided SU-SISO and conventional RIS systems when $N=4$.}\label{fig2:sisoN4}
\end{center}
\end{figure}

\begin{figure}[]
\begin{center}
\subfigure[Optimal elements positions]{\includegraphics[width=\columnwidth]{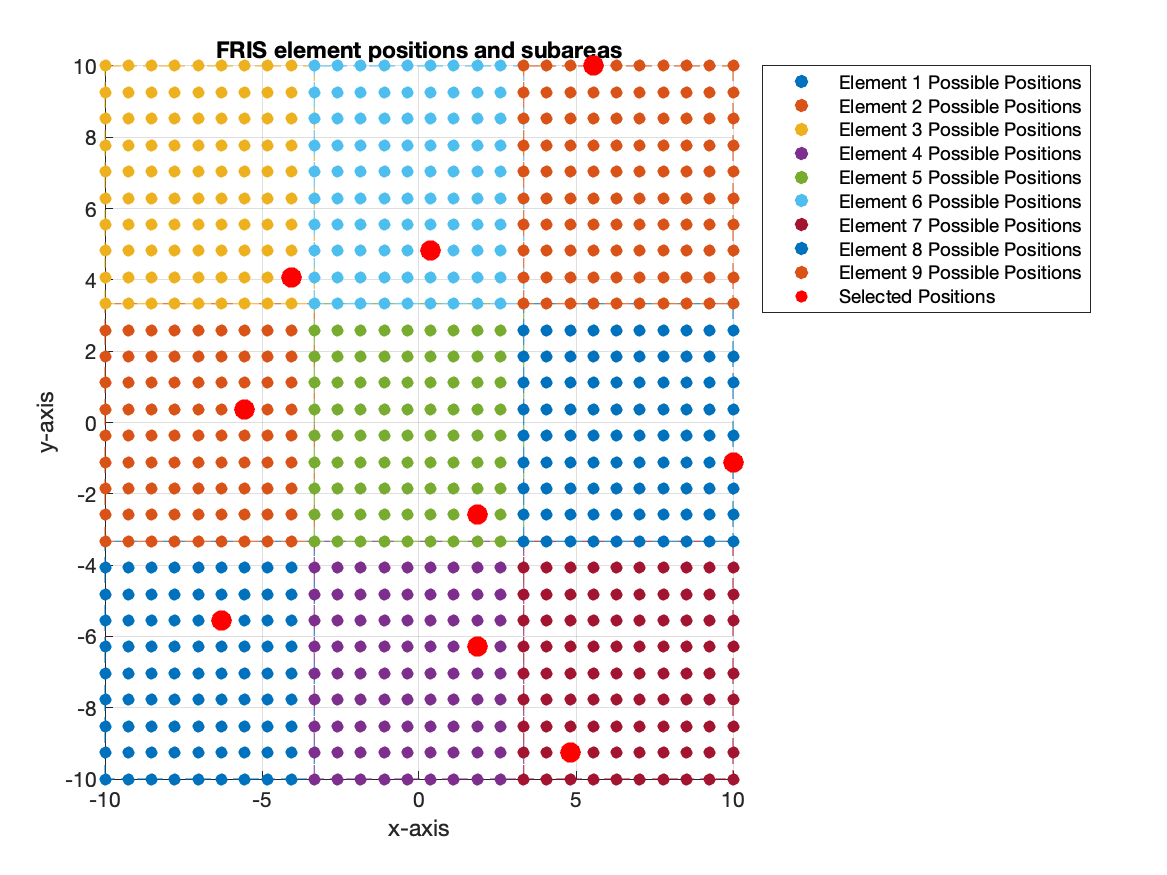}\label{fig3:N9pos}}
\subfigure[Rate performance]{\includegraphics[width=.95\columnwidth]{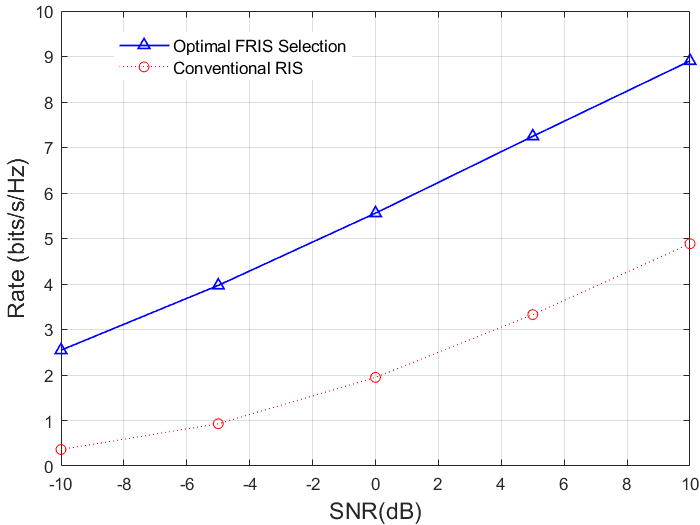}\label{fig3:N9rate}}
\caption{Comparison between the proposed FRIS-aided SU-SISO and conventional RIS systems when $N=9$.}\label{fig3:sisoN9}
\end{center}
\end{figure}

\begin{figure}[]
\begin{center}
\subfigure[Optimal elements positions]{\includegraphics[width=\columnwidth]{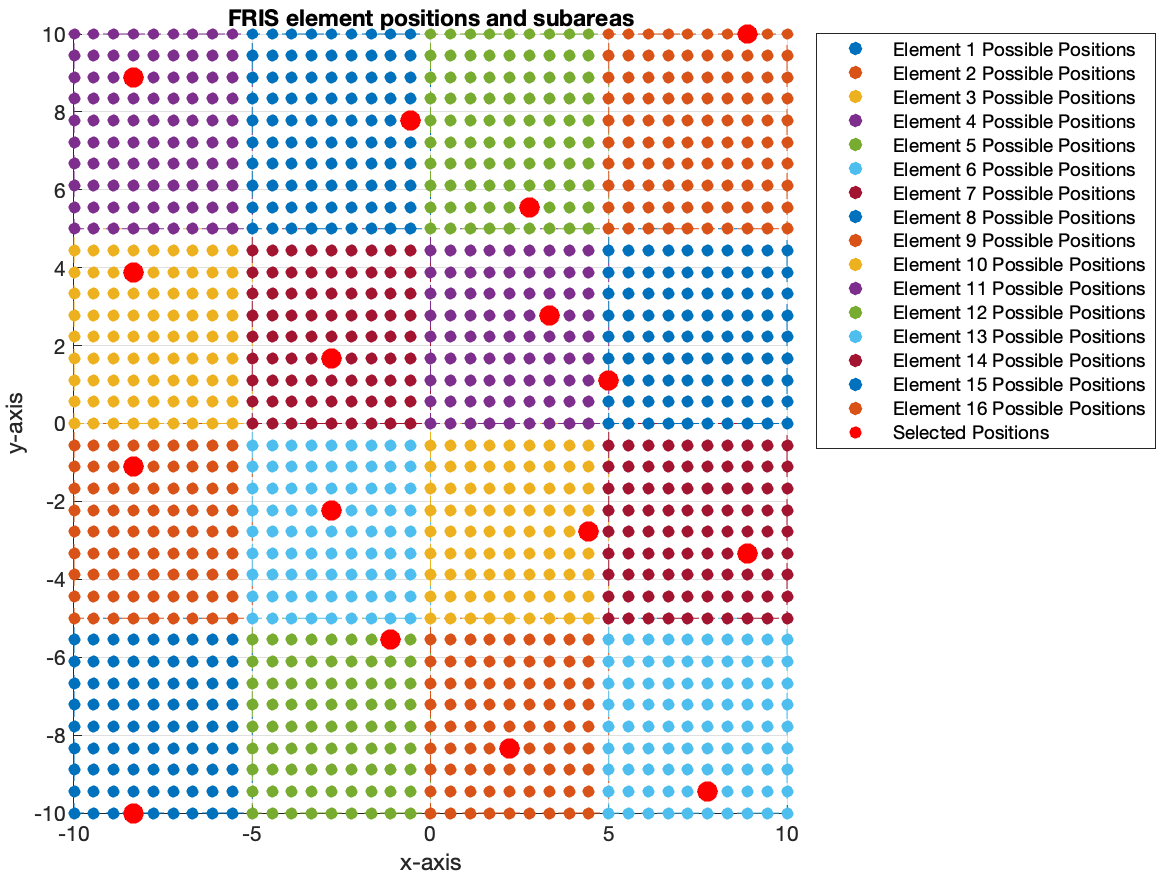}\label{fig4:N16pos}}
\subfigure[Rate performance]{\includegraphics[width=.95\columnwidth]{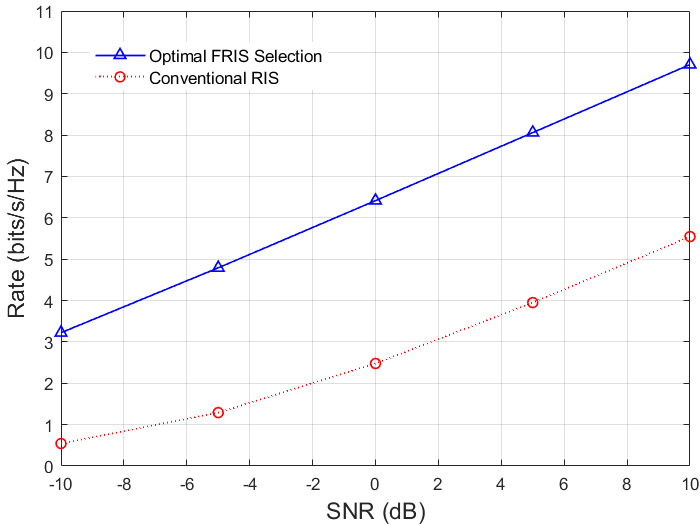}\label{fig4:N16rate}}
\caption{Comparison between the proposed FRIS-aided SU-SISO and conventional RIS systems when $N=16$.}\label{fig4:sisoN16}
\end{center}
\end{figure}

We first examine the performance in the case of SU-SISO systems with the aid of FRIS. The rate performance results are provided in Figs.~\ref{fig2:sisoN4}--\ref{fig4:sisoN16} for $N=4,9,16$, respectively. Fig.~\ref{fig2:N4pos} illustrate the possible fluid element positions on FRIS in the simulations when there are only $N=4$ fluid elements. In that figure, an example of the best fluid element positions is also shown. As can be seen, the optimized positions indicate strategic positioning in the subareas for performance enhancement. Evidently, the exact relationship between the achievable rate and the fluid element positions is complicated and depends on the instantaneous channels.

The results in Figs.~\ref{fig2:N4rate}, \ref{fig3:N9rate} and \ref{fig4:N16rate} demonstrate promising rate performance for using FRIS compared to the conventional RIS systems. As expected, the achievable rates in all the cases increase with the number of elements and the SNR. More importantly, significant rate increase is observed utilizing the proposed FRIS system. Specifically, at $10~{\rm dB}$ of SNR, a rate increase from $4$ bits/s/Hz to $8$ bits/s/Hz is achieved when $N=4$. Similar gains are consistently observed when $N=9$, from $6.5$ bits/s/Hz to $9.5$ bits/s/Hz, and $N=16$, from $7.5$ bits/s/Hz to $10.5$ bits/s/Hz over conventional RIS systems.

\begin{figure}[!t]
\centerline{\includegraphics[width=0.95\columnwidth]{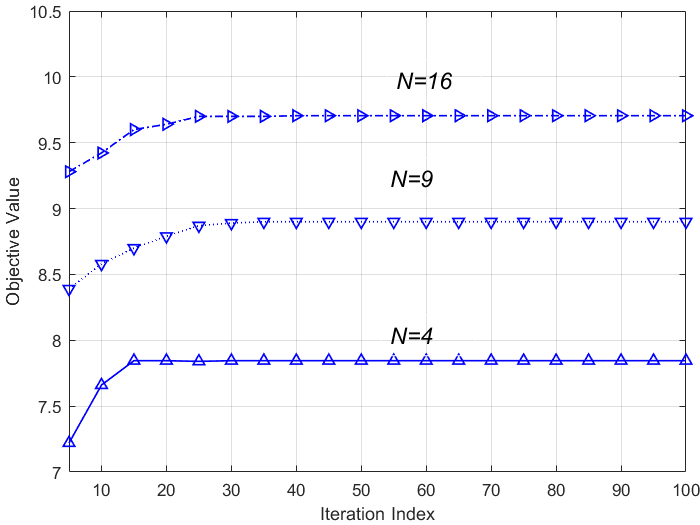}}
\caption{Convergence performance of Algorithm \ref{alg:PSO-siso} when SNR is $10~{\rm dB}$.}\label{fig5:rateVSi}
\end{figure}

In Fig.~\ref{fig5:rateVSi}, we investigate the convergence performance of Algorithm \ref{alg:PSO-siso} for the optimization of fluid element positioning using the PSO-based method. In the simulations, $10~{\rm dB}$ SNR is assumed and different values of $N$ are considered. The results in this figure show that the proposed PSO-based algorithm converges quickly although more iterations would be required if more fluid elements are involved. Decent performance after $10$ or $20$ iterations before convergence can be obtained.

\subsection{MU-MISO}
\begin{figure*}[]
\begin{center}
\subfigure[Rate performance when $N=4$]{\includegraphics[width=.95\columnwidth]{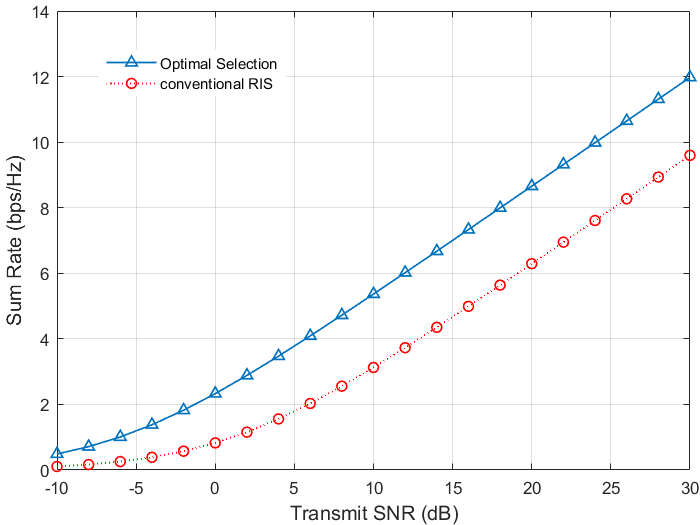}\label{fig6a:N4-miso}}
\subfigure[Rate performance when $N=9$]{\includegraphics[width=.95\columnwidth]{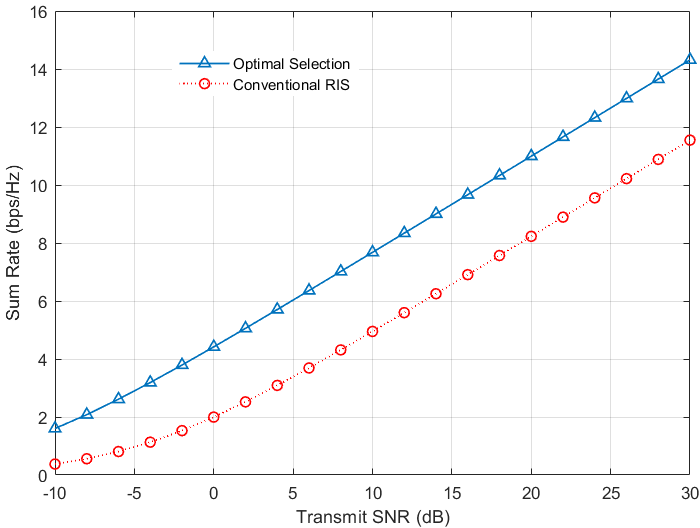}\label{fig6b:N9-miso}}
\subfigure[Rate performance when $N=16$]{\includegraphics[width=.95\columnwidth]{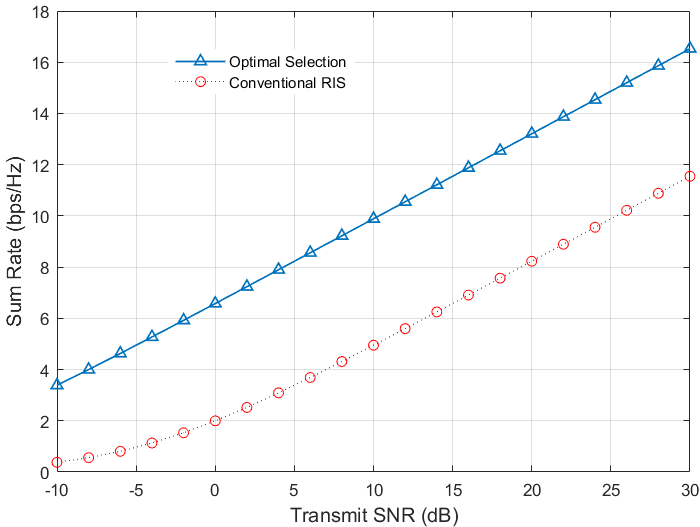}\label{fig6c:N16-miso}}
\caption{Comparison between the proposed FRIS-aided MU-MISO and conventional RIS systems.}\label{fig6:miso}
\end{center}
\end{figure*}

In this subsection, we turn our attention to the MU-MISO setup in which multiple users are considered and the BS has multiple FPAs for precoding. The rate performance results for the FRIS-aided MU-MISO system are provided in Fig.~\ref{fig6:miso} for different values of fluid elements, $N$, while the candidate positions for the fluid elements are the same as the case of SU-SISO before and thus omitted. In the multi-user case, the sum-rate performance is considered, and the BS precoding is optimized together with the fluid element positioning. First of all, it is expected that the rate increases with the number of elements and SNR. Also, FRIS continues to outperform clearly the conventional RIS counterpart, demonstrating the good use of spatial diversity in position reconfigurability of the fluid elements. However, one interesting observation is regarding FRIS with $N=4$ and conventional RIS with $N=16$. At SNR of $30~{\rm dB}$, both systems achieve a rate of about $14$ bits/s/Hz, meaning that FRIS is able to achieve a similar sum-rate performance but with much less number of elements. This also indicates that position reconfigurability of elements can restore the DoF lost by a reduced number of elements, which may have strong practical relevance to RIS technologies.

\begin{figure}
\centerline{\includegraphics[width=0.95\columnwidth]{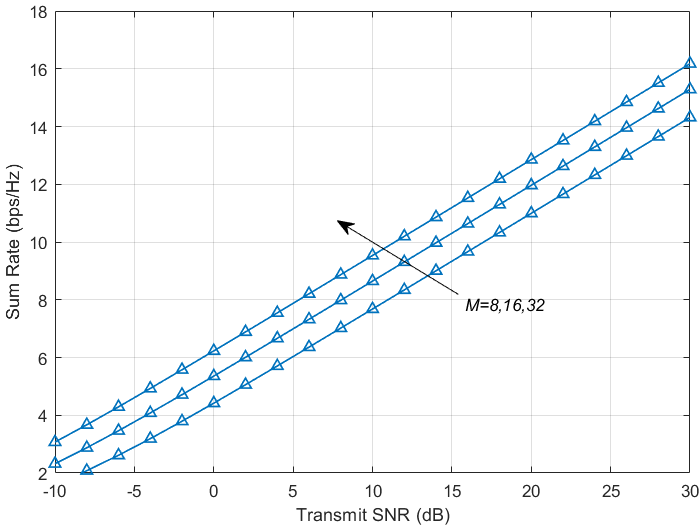}}
\caption{Sum-rate performance of FRIS-aided MU-MISO against the SNR with different number of BS antennas when $N=9$.}\label{fig9:misoVSm}
\end{figure}

Fig.~\ref{fig9:misoVSm} evaluates the impact of the number BS antennas in the sum-rate performance for the FRIS-aided MU-MISO system and we provide the rate results against the SNR for different values of $M=8,16,20$. The results demonstrate that the sum-rate improves as the number of BS antennas increases but it does have a diminishing return. This can be explained by the fact that the sum-rate is dictated by the channel rank which is limited by the fixed number of elements in the FRIS.

\begin{figure}
\centerline{\includegraphics[width=0.95\columnwidth]{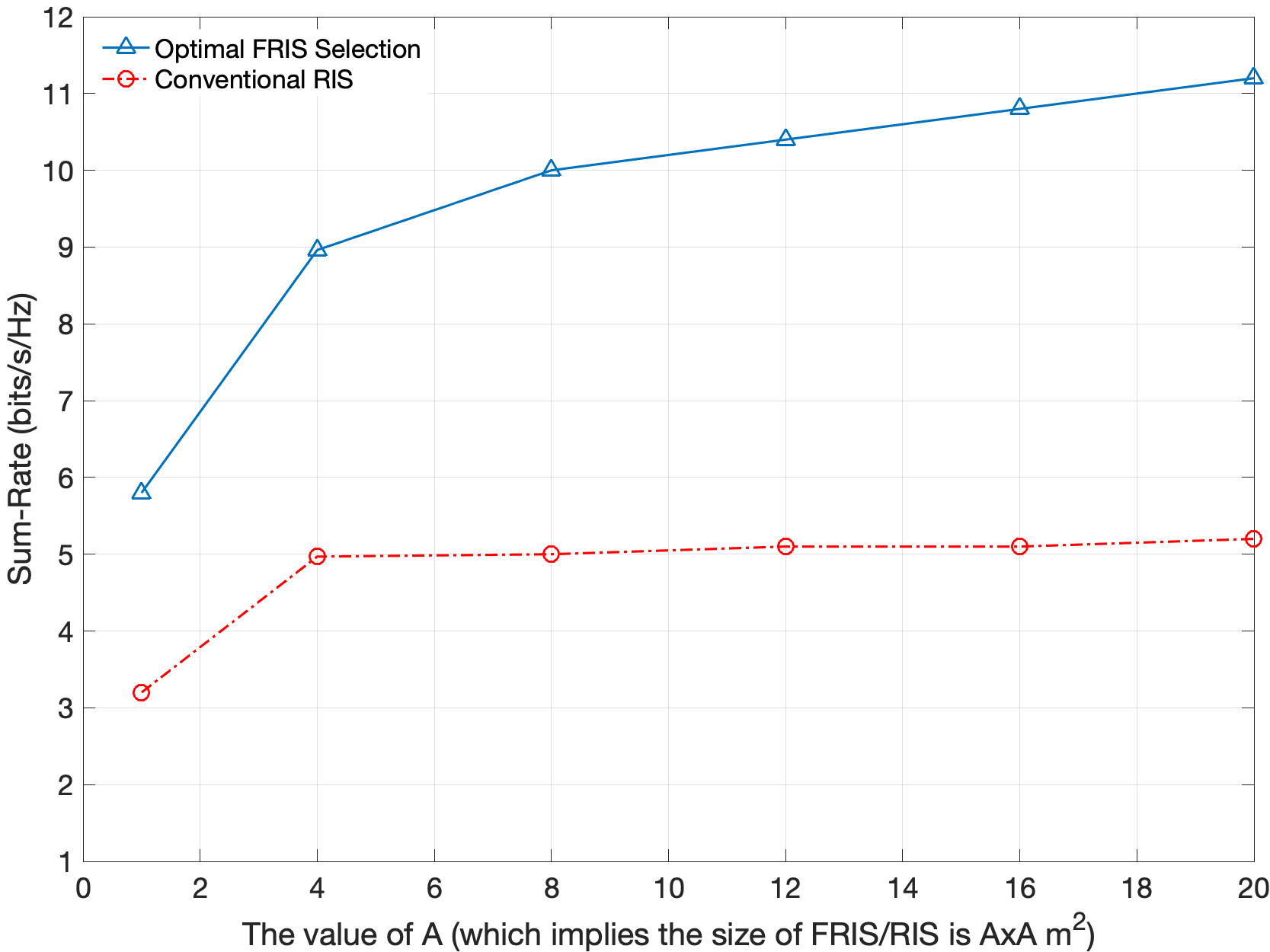}}
\caption{Sum-rate performance of FRIS-aided MU-MISO against the size of FRIS/RIS when ${\rm SNR}=5~{\rm dB}$ and $N=9$.}\label{frisArea}
\end{figure}

Fig.~\ref{frisArea} reveals the impact of the size of FRIS on the sum rate assuming ${\rm SNR}=5~{\rm dB}$ and $N=9$. The results indicate that increasing the area significantly improves the sum rate if FRIS is used while the conventional RIS saturates quickly. This means that even if the number of elements remains the same, there is significant spatial diversity that can be obtained using position reconfigurability. In particular, according to the results in this figure, if the size of FRIS is large enough, it is possible to achieve more than double the sum rate compared to the RIS counterpart, demonstrating the effectiveness of FRIS.

\begin{figure}
\centerline{\includegraphics[width=0.95\columnwidth]{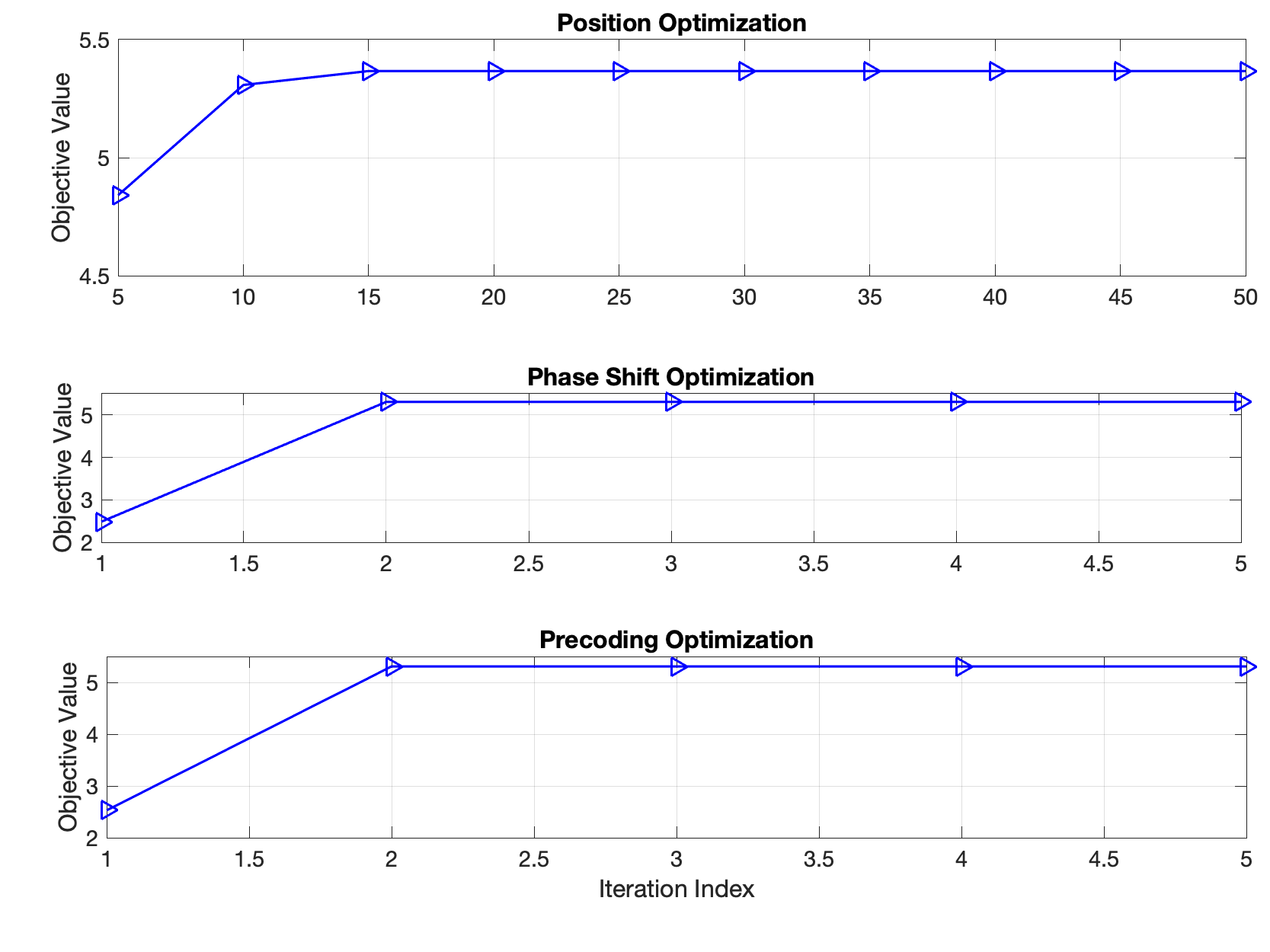}}
\caption{Convergence performance of each sub-problem of Algorithm \ref{alg:W} when there are $N=4$ fluid elements and the SNR is $10~{\rm dB}$.}\label{fig10:subproblem}
\end{figure}

Finally, we use the results in Fig.~\ref{fig10:subproblem} to study the convergence performance for solving each sub-problem in the optimization of the FRIS-aided MU-MISO system. As can be seen, the proposed algorithms converge very quickly, ensuring that the system converges to an optimal configuration with a smaller number of iterations. Specifically, the phase shift and precoding optimization only need two iterations to converge in each round while the position optimization takes $10$ to $15$ iterations to reach a steady state. Overall, these results validate the computational feasibility of the proposed alternating optimization framework for optimizing the FRIS network.

\section{Conclusion\label{sec:conclude}}

This paper proposed the new paradigm of FRIS-aided wireless communications, where each FRIS element is position-reconfigurable, enhancing the DoF for system optimization. We considered both SU-SISO and MU-MISO models and addressed the achievable rate (or sum-rate) maximization problem. For the SU-SISO scenario, we formulated the problem to obtain the optimal fluid element positions and solved it using a PSO-based approach. We then extended the framework to the MU-MISO case, where FRIS element positions, phase shifts, and BS precoding were jointly optimized in an iterative manner to maximize the system sum-rate. The proposed iterative scheme tackled the joint optimization by decomposing it into three subproblems, which were solved using a combination of grid search, PSO, SDR, and MMSE methods. Numerical results demonstrated that FRIS significantly outperforms conventional RIS configurations. Notably, FRIS can achieve the same rate with a much smaller number of elements compared to conventional RISs, highlighting its efficiency and potential for next-generation wireless networks.

\bibliographystyle{IEEEtran}

\end{document}